\documentclass[%
 reprint,
superscriptaddress,
nofootinbib,
 amsmath,amssymb,
 aps,
floatfix,
]{revtex4-2}
\usepackage{graphicx}% Include figure files
\usepackage{dcolumn}% Align table columns on decimal point
\usepackage{bm}% bold math
\usepackage{xcolor}
\usepackage[colorlinks=true,linkcolor=blue,citecolor=blue]{hyperref}% add hypertext capabilities
\usepackage[mathlines]{lineno}% Enable numbering of text and display math
\usepackage{csquotes}
\usepackage{lipsum}
\usepackage[normalem]{ulem}
\usepackage{mathtools}
\usepackage{comment}

\newcommand{\commentGT}[1]{ {\color{green} GT: #1}}
\newcommand{\commentSRS}[1] {{\color{magenta} SRS: #1}}

\begin{document}

\preprint{APS/123-QED}

\title{Ab initio in-medium similarity renormalization group for open-shell atomic systems}% Force line 

\author{G.~Tenkila}
 \affiliation{TRIUMF, Vancouver, BC V6T 2A3, Canada}%

 \affiliation{Department of Physics \& Astronomy, University of British Columbia, Vancouver, British Columbia V6T 1Z1, Canada}

\author{V. Chand}
 \affiliation{TRIUMF, Vancouver, BC V6T 2A3, Canada}%

\author{T.~Miyagi}
 \affiliation{Technische Universit\"at Darmstadt, Department of Physics, 64289 Darmstadt, Germany}
 \affiliation{ExtreMe Matter Institute EMMI, GSI Helmholtzzentrum f\"ur Schwerionenforschung GmbH, 64291 Darmstadt, Germany}
 \affiliation{TRIUMF, Vancouver, BC V6T 2A3, Canada}%
 
\author{H. Patel}
 \affiliation{TRIUMF, Vancouver, BC V6T 2A3, Canada}%
 \affiliation{Department of Physics \& Astronomy, University of British Columbia, Vancouver, British Columbia V6T 1Z1, Canada} 
 
\author{S.~R.~Stroberg}%
\affiliation{Physics Division, Argonne National Laboratory, Lemont, Illinois 60439, USA}

\author{R.~F.~Garcia Ruiz}
\affiliation{Department of Physics, Massachusetts Institute of Technology, Cambridge, MA 02139, USA  }%

\author{J.~D.~Holt}
 \affiliation{TRIUMF, Vancouver, BC V6T 2A3, Canada}%
\affiliation{Department of Physics, McGill University, Montr\'eal, QC H3A 2T8, Canada}

\date{\today}

\begin{abstract}
Precise theoretical calculations of open-shell atomic systems are critical for extracting fundamental physics parameters from precision experiments. Here we present proof-of-principle calculations illustrating the effectiveness of the valence-space formulation of the ab initio in-medium similarity renormalization group, widely used in nuclear theory, as a new approach to atomic systems. We adapt this approach to study properties of closed- and open-shell many-electron systems from helium to calcium. Ground-state energies, excitation spectra, and ionization energies are obtained for light atoms, and reasonable agreement is found with benchmark coupled-cluster and many-body perturbation theory calculations, where available. 

\end{abstract}

\maketitle

%\tableofcontents

\section{\label{sec:introduction}Introduction}

Atomic systems offer unique laboratories in the development of fundamental physics. 
Atomic energy levels can be highly sensitive to nuclear electromagnetic properties~\cite{Cam16,Yan22,Ver22}, and precision measurements of their properties can also provide powerful tests of the Standard Model (SM) at low energy~\cite{Saf18}. 
The resurgence of interest in many-body electron systems is intrinsically linked to known open problems of the standard model (SM). 
Despite the remarkable success of the SM, the absence of answers to fundamental mysteries like dark matter, the hierarchy problems related to the instability of the electroweak scale, and the absence of CP violations in strong interactions highlights its shortcomings. 
As the deficiencies of the SM fail to point towards a new cohesive theoretical hierarchy, it remains important to keep searching for new physics at different energy scales. 
Therefore, as large-scale colliders such as those at CERN explore high-energy physics, it is critical we also probe the low-energy region. 
This has prompted a foundational refurbishment of experimental atomic physics as unique probes of nuclear and particle physics phenomena.

At first order, atomic transitions can be highly sensitive to changes in nuclear density distribution~\cite{Gar16,Mil19,Gor19}. 
However, if the nuclear properties are fully constrained, a precise measurement of atomic energy shifts can reveal subtle details of the electron-nucleus interaction~\cite{Ber18,Sta18,Del17}. 
Recently, the analyses of atomic experiments has spurred developments in uncovering potential long range interactions from electron-neutron coupling and the postulated light bosons~\cite{Saf18,Ber18,Sta18,Dzuba2017,Del17,Del17b,Fru17,Fla18,Hu22}.

Despite the cascade of insight which such studies have launched, the methods are, in many cases, limited by the accuracy of theoretical atomic parameters~\cite{Ant18,Ohayon2022,Sah21}. 
As such this sets a strong precedence for ab initio theory, which aims to understand the properties and structure of atoms from only the underlying interactions between electrons, to provide a methodology to determine atomic parameters with small uncertainties. 
Despite this, there are few ab initio calculations of open-shell atomic systems beyond the Hartree-Fock (HF) mean field level~\cite{Yan2008, Korobov2019, Sumeet2022, Safronova2009}. 
The most successful explorations of ab initio open-shell nuclei have come in the form of coupled cluster and variants of many-body perturbation theory~\cite{Safronova2009, Stopkowicz2015, Porsev2001, Haque1985,Ruiz2018,Sah20}. 
Albeit both fruitful in atomic theory, these methods face considerable limitations when scaling to open-shell systems far from filled atomic orbitals. 
A potential solution arises from the driven similarity renormalization group (DSRG) from quantum chemistry theory~\cite{Li2019}, which is not dissimilar to the approach we introduce here.

In this article we propose the in-medium similarity renormalization group (IMSRG) as a complementary non-perturbative many-body method for atomic systems, where continuous unitary transformations are constructed to evolve the many-body Hamiltonian with a particular chosen generator~\cite{Herg16PR}. 
In particular, the valence-space formulation of the IMSRG (VS-IMSRG) is a broadly applicable extension capable of providing an array of observables related to ground and excited states of essentially all open-shell atoms accessible to the traditional configuration interaction methods~\cite{Herg16PR,Stro17ENO,Stroberg2019}. 
We provide the proof of concept for the applicability of the VS-IMSRG, through comparisons of ground- and excited state properties with existing many-body methods and experimental data for closed and open-shell atomic systems.

\section{\label{sec:theory} Theoretical Approach}

The core problem to solve given an atomic system is diagonalizing the many-body electronic Hamiltonian. 
The IMSRG approach to the problem is to define a flow on the Hamiltonian that continuously evolves it as a function of a flow parameter $s$. 
We can derive a flow equation for the Hamiltonian by considering the action of a unitary operator on it \cite{Wegner1994, Glazek1993}:
\begin{align}
    H(s) = U(s)^\dagger H_0 U(s) \label{eq:0} \\
   \implies  \frac{d}{ds}H(s) = [\eta(s), H(s)] \label{eq:1}
\end{align}
Here, we are redefining the flow in terms of an anti-Hermitian generator $\eta(s)$, and anti-Hermiticity is enforced by the requirement that the flow is unitary. 
The unitarity then ensures that the flow is isospectral, i.e., the Hamiltonian spectrum is an invariant of the flow. We employ the arctan variant of the White generator \cite{White2002, Hergert2013}, given by:
\begin{equation}
    \eta^{\text{atan}}_{ij}(s) = \frac12\text{arctan}\left(\frac{2H^{\text{OD}}_{ij}(s)}{\Delta_{ij}(s)}\right)
\end{equation}
where $H^{\text{OD}}(s)$ is the off-diagonal component of the Hamiltonian and $\Delta_{ij} = H_{ii}(s) - H_{jj}(s)$. 
The arctan regulates divergences that can occur from the energy denominator vanishing.
Other popular choices of generator are the Wegner and the imaginary time generator \cite{Wegner1994, Carlson2015}, where in principle, the diagonalization should be independent of the choice of generator. 
The arctan generator, however, has been shown to be numerically more efficient in nuclear systems, whereas the Wegner generator generally leads to stiff ODEs. In this work, we tested all three generators for atomic systems and found that the numerics is essentially invariant to generator choice. 

In order to systematically define the flow for a given many-body atomic Hamiltonian, we normal order with respect to the HF ground state. 
Assuming weak interactions, the HF ground state is a good approximation to the true ground state of the system. 
Thus the IMSRG flow only needs to account for the electron-electron correlation energies. 
In second-quantized form, the normal-ordered Hamiltonian reads:
\begin{align}\label{eq:2}
     H(s) &= E_{0}(s) + \sum_{ab} f_{ab}(s) \{a^{\dag}_{a} a_{b}\}
     \notag \\ & \hspace{1em}
 + \frac{1}{4} \sum_{abcd} \Gamma_{abcd} \{a^{\dag}_{a}a^{\dag}_{b}a_{d}a_{c}\}
\end{align}
where $E_{0}$, $f_{ab}$, and $\Gamma_{abcd}$ are zero-, one- and two-body matrix elements of Hamiltonian, respectively. 
Evaluating the commutator in Eq.~\ref{eq:1} allows us to derive differential equations for each of the n-body terms.
It is to be noted however that the IMSRG flow induces three-body and higher terms. 
In our calculations, we truncate at the two-body level, which prescribes the so-called IMSRG(2) scheme. 

\begin{figure}[t]
    \centering
    \includegraphics[clip,width=\columnwidth]{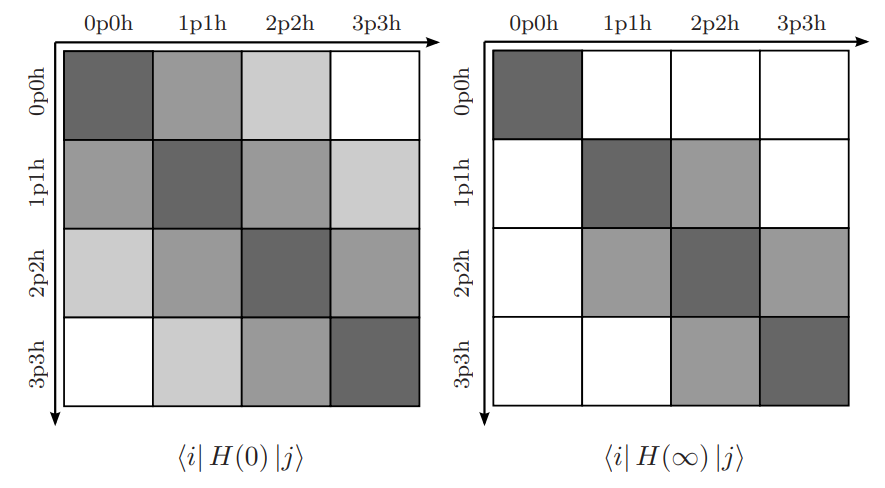}
    \caption{Schematic of IMSRG decoupling of initial and final many-particle Hamiltonian spanned by particle-hole excitations from Hartree Fock reference state \cite{Herg16PR}.}
    \label{fig:block-decouple}
\end{figure}

Since atomic eigenstates cluster into blocks of fixed total angular momentum $J$, it is more practical to block diagonalize the atomic Hamiltonian. 
In particular, we divide the Hilbert space of the system into core, valence and ``outside" subspaces. 
The block diagonalization decouples a valence-space Hamiltonian, which, when exactly diagonalized itself, reproduces eigenvalues of the full-space Hamiltonian. 
Therefore the power of the VS-IMSRG lies in giving systematic access to all open-shell systems \textemdash{} we first decouple the underlying noble atom core then treat valence electrons with the decoupled valence-space Hamiltonian. 

The final theoretical tool we employ is the Magnus expansion \cite{Magnus1954, Morr15Magnus}. 
The IMSRG equations derived from Eq.~\ref{eq:1} and Eq.~\ref{eq:2} can be solved using traditional coupled ODE solvers. 
However, large system sizes lead to memory issue storing the solution vectors, a problem which is compounded if one is required to compute additional observables beyond energies. 
The Magnus expansion addresses this by providing a systematic way to construct the unitary operator that diagonalizes the Hamiltonian. 
Formally, the unitary transformation in Eq.~\ref{eq:0} obeys the flow equation:
\begin{equation}
    \frac{d U(s)}{d s}=-\eta(s) U(s).
\end{equation}
The solution is given by the standard time-ordered exponential of the generator $\eta(s)$. 
The Magnus expansion absorbs the time-ordering and provides a solution of the form:
\begin{equation}
    U(s) = e^{\Omega(s)},
\end{equation}
where the operator $\Omega(s)$ is given by a series expansion:
\begin{gather}
    \Omega=\sum_{n=1}^{\infty} \Omega_n \\
\Omega_1(s) =-\int_0^s d s_1 \eta\left(s_1\right) \\
\Omega_2(s) =\frac{1}{2} \int_0^s d s_1 \int_0^{s_1} d s_2\left[\eta\left(s_1\right), \eta\left(s_2\right)\right] \\
\vdots  \nonumber
\end{gather}
The upshot of the Magnus expansion is that truncating the series still produces a unitary operator. 
Thus, we can diagonalize any commuting observable by explicit construction of the unitary operator. 

\section{\label{sec:analysis} Numerical Analysis}

We assume a point-like nucleus and non-relativistic Coulomb interactions, which gives us the standard Hamiltonian for an atom with atomic number $Z$:
\begin{equation}
    H = \sum_{i=1}^{Z}\left( \frac{p_i^2}{2m_e} - K\frac{Ze^2}{r_i}\right) + \sum_{i<j} \frac{e^2}{r_{ij}},
\end{equation}
where $(\boldsymbol{r}_i, \boldsymbol{p}_i)=(\boldsymbol{r}_i, -i\hbar\boldsymbol{\nabla}_i)$ are the canonical position and momentum operators for the $i$-th electron, with $r_{ij} = |\vec{r}_i-\vec{r}_j|$, while $m_e$ and $e$ are the electron rest mass and charge respectively. 
For numerical simplicity, we work in atomic Hartree units, which sets the numerical values $e=m_e=a_0=\hbar=1$ ($a_0$ is the Bohr radius). 
This also sets Coulomb's constant $K=(4\pi\epsilon_0)^{-1}$ to unity.
A choice of an orthogonal and complete basis set $\{|{\phi_i}\rangle\}$ is required for the matrix representation of the Hamiltonian $H_{ij} = \langle \phi_i|H| \phi_j\rangle$. 
A standard choice made for nuclear systems is the harmonic oscillator (HO) basis. 
However, we found that this set has poor convergence properties for atomic systems because of the Gaussian fall in the asymptotic region. 
The optimal basis set was instead found to be the Laguerre orbital Basis, which has the form \cite{McCoy2016}:
\begin{equation}
    \Lambda_{n l m}(\mathbf{r}) \propto( 2r{\zeta})^l L_n^{2 l+2}( 2r{\zeta}) e^{-r\zeta} Y_{l m}(\hat{\mathbf{r}})
\end{equation}
where $L^{2l+2}_n$ are the generalized Laguerre polynomials and $Y_{lm}(\hat{\mathbf{r}})$ are the usual spherical harmonic functions. 
The $n, \ l,$ and $ m$ indices correspond to the principle, angular momentum, and magnetic quantum numbers respectively, where $\zeta$ is an inverse length scale we introduce to tune the basis set.
The basis has a similar form to the Coulomb-Sturmian functions~\cite{Hylleraas1928, ROTENBERG1962262}, thus making it ideal for representing bound state electronic wave-functions, exponential asymptotic behavior. 

We use two independent parameters to determine the optimal truncation for the single-particle basis.
The first is termed $e_{\mathrm{max}}$ and is defined as $e_{\mathrm{max}} \coloneqq n + l$ and determines the truncation of the basis set used. 
The second is the inverse length scale $\zeta$, which physically corresponds to the radial width of the orbitals and roughly scales as the atomic number. 
We also use truncation in the angular quantum number $l$ for generating the Hamiltonian matrix elements, denoting it as $l_{\rm max}$. For calculations presented here, we use a global $l_{\rm max} = 10$ since our atoms of interest fill only $s$- and $p$-orbitals.
The numerical IMSRG decoupling and subsequent configuration-interaction calculation were done with the imsrg++~\cite{imsrg++} and KSHELL~\cite{Shimizu2019} codes, respectively.

\begin{figure}[t]
    \centering
    \includegraphics[clip,width=\columnwidth]{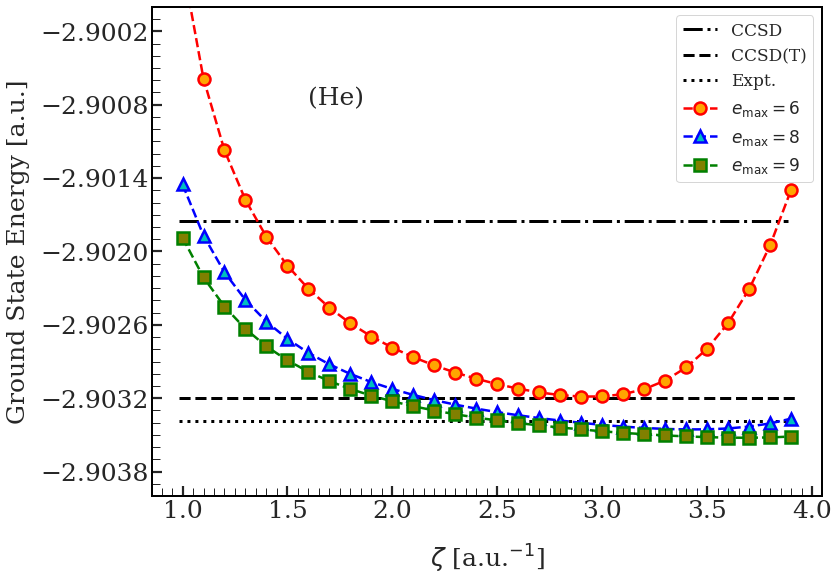}
    \caption{Convergence of helium ground-state energy for $e_{\mathrm{max}}=6,8$ within the IMSRG. The dotted lines depict the experimental value~\cite{NIST_ASD}, while dashed/dot-dashed lines depict values computed with coupled-cluster theory at the CCSD and CCSD(T) levels, respectively~\cite{Stopkowicz2015}. In comparison, the Hartree-Fock ground state energy is $E_{\mathrm{HF}}=-2.8616 \ \mathrm{a.u.}$ The shift in optimal $\zeta$ is due to different rates of convergence in the UV and IR regimes.} 
    \label{fig:He_gs}
\end{figure}

\begin{figure}[t]
    \centering
    \includegraphics[clip,width=\columnwidth]{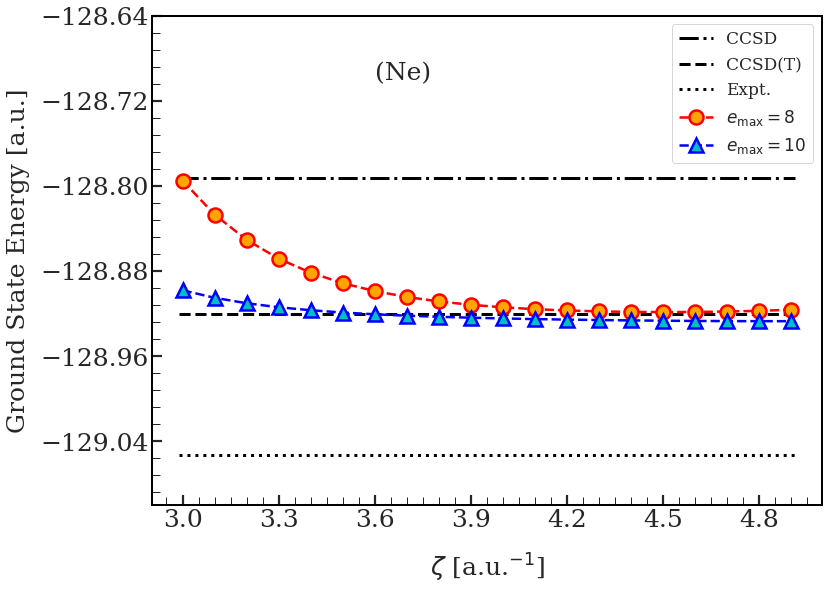}
    \caption{ Convergence of neon ground-state energy for $e_{\mathrm{max}}=8,10$ within the IMSRG. Dotted lines depict the experimental value~\cite{NIST_ASD}, and dashed/dot-dashed lines depict values computed with coupled-cluster theory at the CCSD and CCSD(T) levels, respectively~\cite{Stopkowicz2015}. Converged IMSRG ground state energy coincides with CCSD computed value of the neon ground state. Discrepancy with experiment can be attributed to fine structure corrections to the Hamiltonian. In comparison, the Hartree-Fock ground state energy is $E_{\mathrm{HF}}=-128.5228 \ \mathrm{a.u.}$ 
    }
    \label{fig:Ne_gs}
\end{figure}

\section{\label{sec:applications} Results and Applications}

\subsection{\label{sec:noble} Closed-Shell Atoms}

Our first investigations center on computing ground-state energies for the noble elements. 
The absence of an occupied valence shell trivializes part of the VS-IMSRG calculation, allowing us to get an independent handle on the IMSRG flow and convergence properties. 
Further, it allows us to compare our results with coupled-cluster calculations, which are optimal for closed-shell and closed-subshell configurations. 
Figures~\ref{fig:He_gs} and \ref{fig:Ne_gs} show the ground-state energies for helium and neon with respect to the basis scaling parameter $\zeta$ and the basis truncation $e_{\mathrm{max}}$. 
The data demonstrates the convergence of the calculations and also is in agreement with the coupled-cluster calculations at the single, double, and perturbative triple [CCSD(T)] level~\cite{Stopkowicz2015}, which we expect from known results for nuclear systems. 

\begin{comment}
\begin{figure}[t]
    \centering
    \includegraphics[clip,width=\columnwidth]{Newer_Plots/He_extended.png}
    \caption{ Raw plots for Helium GSEs}
    \label{fig:He_gs_ext_raw}
\end{figure}
\begin{figure}[t]
    \centering
    \includegraphics[clip,width=\columnwidth]{Newer_Plots/He_extended_clean.png}
    \caption{ Cleaned up a bit and stitched together. I'm not sure what to make of the $\zeta$ dependence, its a bit strange.
    \commentSRS{Strange compared to what? It looks like the left side is dominated by UV convergence and the right by IR convergence. The optimal $\zeta$ shifts higher with bigger emax, maybe because the IR converges faster\commentGT{Okay yeah that makes sense.}}
    }
    \label{fig:He_gs_raw}
\end{figure}
\end{comment}

We note, however, that there is a larger discrepancy between the computed and experimental values for the ground state of neon, compared to helium. 
This is an expected effect, since we are assuming a point-like nucleus with a non-relativistic Coulomb potential. 
As the atomic number increases, the nuclear structure and relativistic corrections offer non-trivial contributions to the spectrum. 
A simple calculation shows that the discrepancy is indeed on the order of the fine-structure constant ($\alpha \approx 1/137$). 

\begin{comment}
    
\begin{figure}[t]
    \centering
    \includegraphics[clip,width=\columnwidth]{Newer_Plots/Be_spec.png}
    \caption{VS-IMSRG converged excitation spectrum for atomic beryllium in comparison with experimental data from NIST \cite{NIST_ASD}. The spectroscopic notation labels are for the experimentally measured values.\commentSRS{It would be good to have the spectrum without IMSRG decoupling for comparison, to show we're actually getting something for those cpu cycles.}}
    \label{fig:Be_spec}
\end{figure}
\end{comment}
\begin{figure}[t]
    \centering
    \includegraphics[clip,width=\columnwidth]{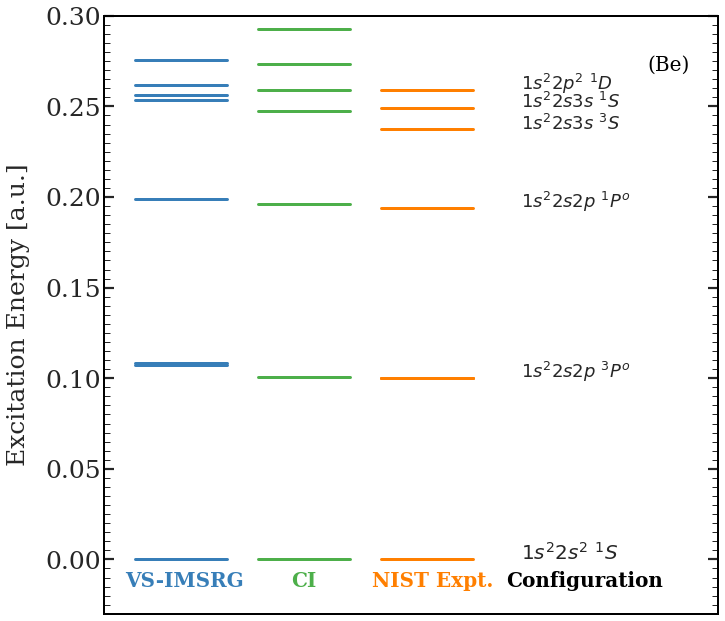}
    \caption{VS-IMSRG converged excitation spectrum for atomic beryllium in comparison with experimental data from NIST \cite{NIST_ASD}. The spectroscopic notation labels are for the experimentally measured values. The green bars in the center are from theoretical computations using the full Configuration Interaction method \cite{Weiss1995}.
    %\commentGT{I found some hyper-optimzed CI calculations through the NIST \cite{Weiss1995} . Not sure whether to include this or not. }
    %\commentSRS{It's fine, but it's also helpful to show a cheaper calculation to show that we're actually improving over something.}
    %\commentGT{Unfortunately, I don't think we have any HF level calculations of this. I found some moderately obscure literature \cite{sanu2018} on variational calculations for something they called "confined beryllium atoms", but the energies reported didn't quite make sense. Edit: The confinement is referring to an infinite potential barrier at some fixed radius. They compute the first excited state in the zero-confining potential limit, and the energy is indeed worse than IMSRG. I don't know if we want to cite this though}
    %\commentSRS{I'm thinking just do the same calculation but don't integrate the flow equations, i.e. set $s_{max}=0$. This should be a relatively cheap calculation.}\commentGT{Maybe Hrishi can work on this? I don't have the code setup on my device and I'm a bit rusty...}\commentJH{Of course this is good to have, but perhaps we could easily add this in a revision? And could you add Be to the figure somewhere?}\commentGT{Updated!}
    }
    \label{fig:Be_spec}
\end{figure}

\subsection{\label{sec:ex_en} Excited States and Ionization Energies}

We can compute excitation energies in the VS-IMSRG framework by exact diagonalization of the effective valence-space Hamiltonian obtained from the renormalization group flow. 
The number of excited states that can be computed is of course limited by the size of the valence space. 
For the atomic systems discussed here, we specify the valence space to be the valence $s-$ and $p-$subshells. 
In Fig.~\ref{fig:Be_spec} we show the energy spectrum of beryllium $Z=4$, compared with data from the National Institute of Standards and Technology (NIST) database \cite{NIST_ASD}. 
We observe overall good agreement with the experimental data, where discrepancies can be attributed to missing fine-structure corrections. 
We also observe that the excited state energies converge at a slower rate than the ground state energies. 
This is to be expected as the excited states have a larger quantum number and would require more terms in the basis expansion.

\begin{figure}[t]
    \centering
    \includegraphics[clip,width=0.9\columnwidth]{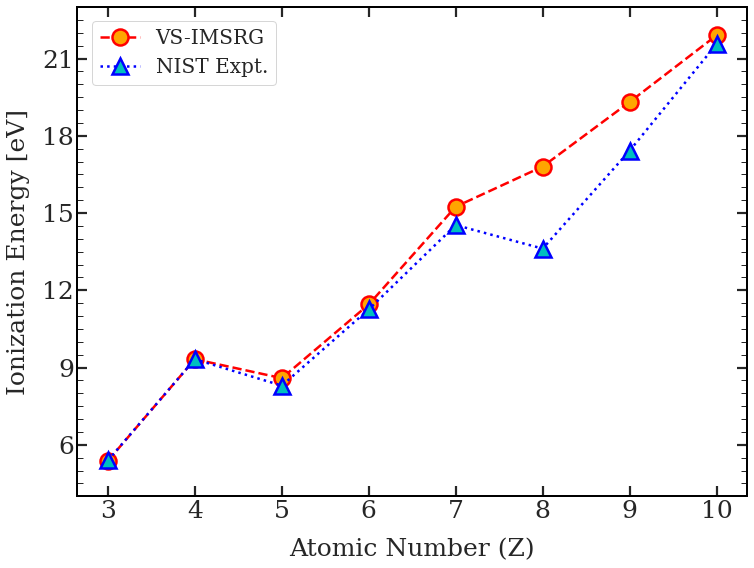}
    \caption{Converged ionization energies for period 2 elements. The VS-IMSRG values (orange circles) are compared with experimental values (blue triangles) \cite{NIST_ASD}}
    \label{fig:p2_ions}
\end{figure}

\begin{figure}[t]
    \centering
    \includegraphics[clip,width=\columnwidth]{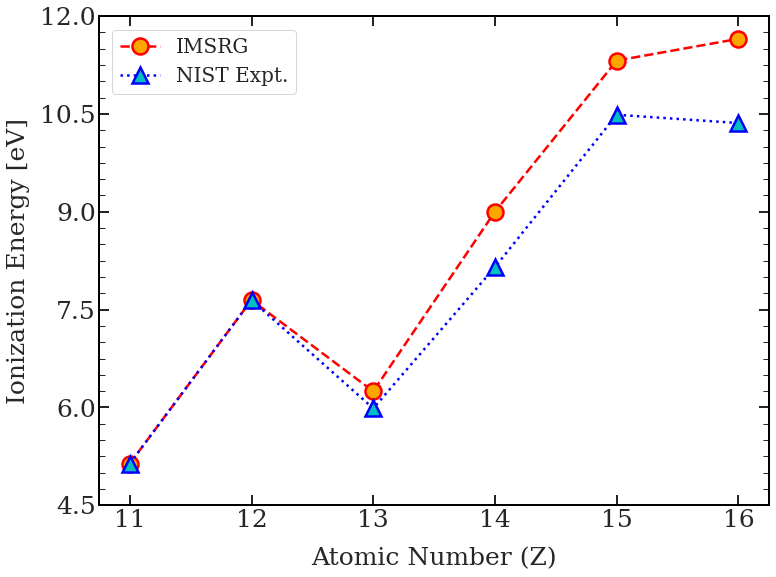}
    \caption{Converged ionization energies for period 3 elements. The VS-IMSRG values (orange circles) are compared with experimental values (blue triangles) \cite{NIST_ASD}
    }
    \label{fig:p3_ions}
\end{figure}

The single-electron ionization energies of atoms are experimentally measured to high accuracy and can be used as a general benchmark for the VS-IMSRG calculations. 
The ionization energy is computed by simply taking the difference in ground-state energies between the neutral atom and a singly ionized cation. 
We systematically performed this calculation up to $Z=16$, as preliminary computations for fluorine and argon ($Z=17,\ 18$) show a large discrepancy with experiment due to growing relativistic corrections. 
This is further exacerbated by issues with convergence given the memory limitations for storing the Hamiltonian elements.
%\commentSRS{Why not to $Z=18$ to finish out period 3?}\commentGT{We have some very early data on Argon when we didn't have fine variation of the $zeta$ param, but the discrepancy with expt. is pretty huge, and we were at the maximum emax that could be handled given memory restrictions. We decided to hold off on going higher till we figured out how to systematically add in relativistic corrections.}
%\commentSRS{Ok. I think something along these lines should be mentioned.}
Figures~\ref{fig:p2_ions} and \ref{fig:p3_ions} show converged VS-IMSRG ionization energies for period 2 and 3 elements, respectively, as compared with NIST experimental data. 
We observe good agreement with experiment for the first four groups, while deviations become prominent past the chalcogens. 
Observing that largest deviation is for oxygen, we hypothesize that the discrepancy could be due to fine structure corrections enhanced by the electron pairing, but further investigation is clearly warranted, as possible deficiencies in the VS-IMSRG could also play a role.

%\commentSRS{Is this really all we have to say about this??? The disagreement definitely seems to be an IMSRG effect, and I'm not an atomic theorist, but if I recall correctly, by Hund's rule up to nitrogen $Z=7$, the orbitals are singly filled and oxygen is the first with a doubly-filled (i.e. $\uparrow \downarrow$ $p$ orbit). Are we using ensemble normal ordering for these calculations? We're also using a $j$ coupled basis, which makes sense for nuclei, but is maybe not ideal for atoms. Is the same reference state used for the neutral atom and the ion?}\commentGT{I did do runs with ENO turned on and off and didn't make an appreciable difference. Yeah this one sorta stumped us - I think what we settled on was with the increasing number of electrons in the valence space, orbit-orbit, spin-orbit interactions become appreciable and these are effects we aren't accounting for. But I really wanted to check if that reasoning makes sense with an atomic person before making that statement. I do remember Ronald mentioning that F and O are particularly difficult to do, so I don't think its entirely unreasonable.}
%\commentSRS{So your suggestion is that the effect is a defect of the Hamiltonian, not of the IMSRG? That seems surprising to me.}
%\commentGT{I think so, yeah. There's a small drop in the energy, which can be accounted for by the sub-shell half closing. I would expect a fairly large spin-spin interaction coming from the electron pairing lowering the ionization energy. I'll try to make this a bit more concrete.}

\begin{figure}[t]
    \centering
    \includegraphics[clip,width=\columnwidth]{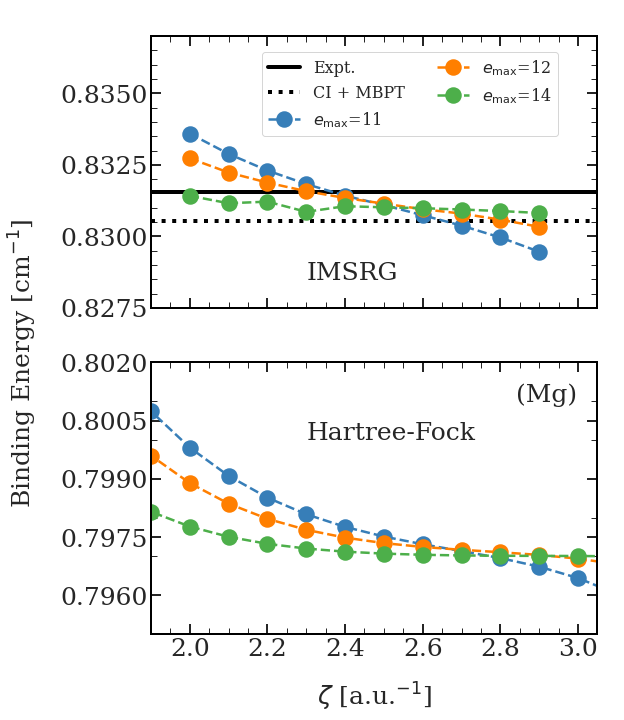}
    \caption{Two electron binding energies for magnesium, benchmarked against CI+MBPT calculations \cite{Safronova2009}.
    }
    \label{fig:Mg_2ebe}
\end{figure}

\begin{figure}[t]
    \centering
    \includegraphics[clip,width=\columnwidth]{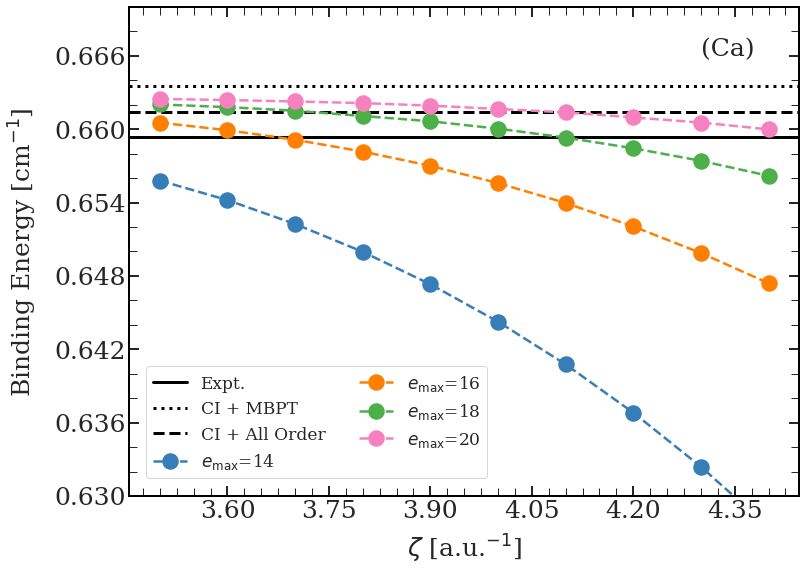}
    \caption{Two electron binding energies for calcium, benchmarked against CI+MBPT calculations \cite{Safronova2009}. %\commentSRS{What basis set is used in the CI+ calculations?} \commentGT{From the Safronova paper: ``The finite basis set of 245 orbitals that include l= 0, ... , 5 partial waves is formed in the spherical cavity with a 50 a.u. radius." Not quite sure what this corresponds to.} \commentSRS{I agree it's not clear how to get 245 from what they say. Also, it looks like the optimal $\zeta$ is lower. Why did you stop at 3.5?} \commentGT{The IMSRG gets a bit finnicky as you approach the d-shell atoms since the level spacings are pretty close, it doesn't get the order right. In fact, its highly dependent on choice of $\zeta$. I did some analysis on this a while ago when I was trying to diagnose a different problem, and I found there's these discrete jumps in the energy that happen at low emax as I vary the $\zeta$. Figure below.}
}
    \label{fig:Ca_2ebe}
\end{figure}

Current state-of-the-art methods for ab initio atomic calculations include the CI+MBPT formalism developed in \cite{Porsev2001}. 
This involves both a non-perturbative and perturbative treatment of the atomic system using configuration interaction (CI) derived ground state and many-body perturbative corrections (MBPT) respectively. 
As a further test of the VS-IMSRG calculations, we compare two-electron binding energies of magnesium and calcium calculated from CI+MBPT, as well as its improved variant CI+All Order~\cite{Safronova2009}, which includes particular diagrams to all orders in perturbation theory. 
The converged binding energies for these two atoms are shown in Figs~\ref{fig:Mg_2ebe} and \ref{fig:Ca_2ebe} respectively. 
We find that the converged VS-IMSRG calculations are in close agreement with the CI techniques and the experimentally measured results, within the order of correlation energies.

Finally, we note that the both relativistic (CI+MBPT) and non-relativistic calculations (IMSRG) yield comparable results for the binding energies, suggesting that the contribution of such effects to the ionization energies is minimal \cite{Safronova2009}. 

\begin{comment}

\begin{figure}[t]
    \centering
    \includegraphics[clip,width=\columnwidth]{Newer_Plots/Ca_kinetic_e.png}
    \caption{Raw plots of kinetic energy of Ca ground state, with varying zeta and emax(x-axis). Potential energy shows similar jumps at lower $\zeta$}
    \label{fig:Ca_KE_raw}
\end{figure}
\end{comment}

\section{\label{sec:summary} Conclusion and Outlook}
In this article, we have introduced the VS-IMSRG as a new tool to provide accurate calculations of open-shell atomic systems, comparable in accuracy with other state-of-the-art ab initio methods such as coupled-cluster theory.
We demonstrate the effectiveness of the VS-IMSRG through several benchmark calculations. 
We first address ground-state energies of light noble atoms with closed-shell structures. We find good agreement with experimentally measured values, noting discrepancies consistent with relativistic corrections. Proceeding to open-shell systems, we systematically compute ionization energies up to $Z=16$, again finding agreement with measured values. 
We also demonstrate that IMSRG is competitive with currently used methods used for atomic structure calculations, namely coupled cluster and CI + MBPT. 
It is clear from both ground-state and excitation energies that fine structure contributions become relevant for elements beyond period 2, especially for mid-period elements. Current work is being done to move to heavier atoms by systematically adding fine structure corrections through a combination of Dirac-Hartree-Fock (DHF) with electron-correlations derived from IMSRG. 
We then aim to eventually implement the DHF ground state as the reference state for the IMSRG flow and perform fully relativistic calculations to address properties of open-shell atoms across the periodic table, including isotope shifts relevant for laser-spectroscopy experiments.

\begin{acknowledgments}
We thank M. Safronova for useful discussions and comments on the manuscript.
TRIUMF receives funding via a contribution through the National Research Council of Canada. 
This  work was further supported by NSERC under grants SAPIN-2018-00027 and RGPAS-2018-52245, the Arthur B. McDonald Canadian Astroparticle Physics Research Institute, the Canadian Institute for Nuclear Physics, the Deutsche Forschungsgemeinschaft (DFG, German Research Foundation) -- Project-ID 279384907 -- SFB 1245), the National Science Foundation under award number 2207996.
Computations were performed with an allocation of computing resources on Cedar at WestGrid and Compute Canada.
\end{acknowledgments}

\bibliography{atomic_bib}

\end{document}